\documentclass[preprint,aps]{revtex4}
\usepackage{graphicx}
\usepackage{dcolumn}
\usepackage{bm}
\begin{document}


\title{Structural, electronic, and optical properties  of ZrO$_2$
 from {\it {ab initio}} calculations}

\author{J. C. Garcia,  L. M. R. Scolfaro}

\address{Instituto de
F$\acute{\imath}$sica, Universidade de S\~{a}o Paulo, CP 66318,
05315-970 S\~{a}o Paulo, SP, Brazil}

\author{A. T. Lino}

\address{Instituto de
F$\acute{\imath}$sica, Universidade Federal de Uberl\^andia, CP
593, 38400-902 Uberl\^andia, MG, Brazil}

\author{V. N. Freire, G. A. Farias}

\address{Departamento de F\'{\i}sica, Universidade Federal do Cear\'a,
CP6030, 60455-900 Fortaleza, CE, Brazil}

\author{C. C. Silva, H. W. Leite Alves}

\address{Departamento de Ci\^encias Naturais, Universidade
Federal de S\~ao Jo\~ao del Rei, CP110, 36301-160 S\~ao Jo\~ao del
Rei, MG, Brazil}

\author{S. C. P. Rodrigues, E. F. da Silva Jr.}

\address{Departamento de F\'{\i}sica, Universidade Federal de
Pernambuco, 50670-901 Recife, PE, Brazil}

\date{September 28, 2006}

\begin{abstract}

Structural, electronic, and optical properties for the cubic,
tetragonal, and monoclinic crystalline phases of ZrO$_2$, as
derived from {\it {ab initio}} full-relativistic  calculations,
are presented. The electronic structure calculations were carried
out by means of the all-electron full potential linear augmented
plane wave method, within the framework of the density functional
theory and the local density approximation. The calculated carrier
effective masses are shown to be highly anisotropic. The results
obtained for the real and imaginary parts of the dielectric
function, the reflectivity, and the refraction index, show good
agreement with the available experimental results. In order to
obtain the static dielectric constant of ZrO$_2$, we added to the
electronic part, the optical phonons contribution, which leads to
values of $\epsilon_{1}(0)\simeq 29.5, 26.2, 21.9$, respectively
along the $xx, yy$, and $zz$ directions,  for the monoclinic
phase, in excellent accordance with experiment. Relativistic
effects, including the spin-orbit interaction, are demonstrated to
be important for a better evaluation of the effective mass values,
and in the detailed structure of the frequency dependent complex
dielectric function.
\end{abstract}

\maketitle

\section{Introduction}

Zirconia (ZrO$_2$) is a material of great technological potential
importance due to its outstanding mechanical and electrical
properties, high dielectric constant and wide band gap. Among the
ZrO$_2$ applications, are: its use as gas sensors, solid fuel
cells, high durability coating, catalytic agents, etc. In recent
years, its gap ($E_g \sim {6}$ eV)~ and dielectric properties
($\varepsilon \sim 25$)~ suggested its potential to replace
SiO$_2$ in advanced metal oxide semiconductor devices (MOS) in
gate stack, dynamic access memory devices, and optical
applications \cite{IRS2004,WilkI,Houssa,Wang,Lin}. Moreover, it
has a large band-offset in direct contact with Si and good thermal
stability. These attracting properties has led zirconia based
oxides to be widely studied in recent years
\cite{JpMaria,WilkII,Nature}.

ZrO$_2$ presents polymorphism with monoclinic, tetragonal, and
cubic phases. The monoclinic phase (m-ZrO$_2$) of undoped ZrO$_2$
is thermodynamically stable at temperatures below 1170 $ ^{\circ}
C$. The tetragonal phase structure (t-ZrO$_2$) arises when heating
ZrO$_2$, being stable between 1170 $ ^{\circ} C$ and 2370 $
^{\circ} C$, as well as above this; through its melting point
(2706 $ ^{\circ} C$), the cubic phase (c-ZrO$_2$) is then
observed. The temperature in which the tetragonal to cubic
transformation occurs can be lowered, by the addition of solutes
such as MgO, CaO, Y$_2$O$_3$, allowing the achievement of the
stabilized c-phase even at room temperature.

Previous investigations of ZrO$_2$, based on first-principles
electronic structure calculations,  have been reported in the
literature
\cite{Lowther,Orlando,Jansen,Kralik,Stefanovich,Zhao,Walter,French}.
Most of them dealt with the  structural and electronic  properties
of  ZrO$_2$, in its cubic, tetragonal, and monoclinic phases
\cite{Lowther,Orlando,Jansen,Kralik,Stefanovich,Walter}. The
zirconia  optical properties, or more specifically its  complex
dielectric function  has been addressed by French {\it et al}
(electronic frequency regime only) by carrying out  {\it ab
initio} electronic structure calculations via the orthogonalized
linear-combination-of-atomic-orbitals method \cite{French}, and
more recently, by Zhao and Vanderbilt \cite{Zhao}  and by
Rignanese {\it et al} \cite{rignanese04}, by means of {\it ab
initio} pseudopotential calculations. However, in the work of
Ref.\cite{rignanese04} only the ZrO$_2$ cubic and tetragonal
structural phases have been considered, while in that of
Ref.\cite{Zhao}, although in the work all three  phases (cubic,
tetragonal, and monoclinic) are studied, only the lattice
contributions to the dielectric tensor have been addressed so far.
Moreover, in these
 studies, relativistic effects including the spin-orbit interaction were not taken into
account in the calculations, neither have been provided values for
the ZrO$_2$ carrier effective masses, which are fundamental to
model the tunnelling currents through floating gate memory devices
that use, e.g., ZrO$_2$ as the tunnelling oxide.

In this work, we present results of state-of-the-art
first-principles calculations for the structural, electronic, and
optical properties of ZrO$_2$ in its monoclinic, tetragonal, and
cubic  phases. From the band structures, the ZrO$_2$ conduction-
and valence-band effective masses have been obtained, in the most
important symmetry axis of the Brillouin zone. The
full-relativistic effects, properly taken into account in the
calculations, are demonstrated to be relevant for a more precise
evaluation of its electron- and hole-effective-mass values, as
well as in the detailed structure of the electronic contribution
to its frequency dependent dielectric function,
($\varepsilon^{elect}(w)$). As a consequence of the role played by
the inclusion of full-relativistic effects in the calculations,
the obtained changes in the reflectivity spectrum and refraction
index are also significant.

In order to realistically obtain the value for the static
dielectric constant $\varepsilon_{1}(0)$, we also calculated  the
lattice contributions, mostly arising from the transverse optical
phonon modes, to the low-frequency dielectric function, for the
different phases. The resulting calculated static dielectric
constant for m-ZrO$_2$ is in excellent agreement with the values
reported in the literature. Moreover, the results obtained for the
reflectivity and refraction index for its monoclinic phase are in
good accordance with the available experimental data.

\section{Calculation Details}

The electronic structure calculations were carried out using the
density-functional theory in the local-density approach (DFT-LDA)
by means of the {\it ab initio} full-potential self-consistent
linear augmented plane wave (FLAPW) method (Wien2k code
\cite{wien2k}). The generalized gradient approximation (GGA) was
adopted for the exchange-correlation potential \cite{GGA}.  A
similar approach has been recently applied to study the
structural, electronic, and optical properties of cubic SrTiO$_3$
and HfO$_2$ \cite{apl-srtio3,apl-hfo2}. The Zirconium
4s-,4p-,4d-,and 5s, and the Oxigen 2s,2p electrons were treated as
part of the valence states. The cutoff angular momentum was $l=10$
for wave functions, and $l=5$ for charge densities and potentials
inside the spheres. Muffin-tin sphere radii were assumed equal for
both Zr and O. The number of symmetrized {\bf k}-points, used as
input for the self-consistent charge density determination, was
47, 75, and 100-{\bf k} points in the irreducible symmetry wedge
of the BZ, respectively for the cubic, tetragonal, and monoclinic
phases. The value $RKMAX=9$ was adopted for the three different
phases. With these assumptions, the energy bands were converged
within $10^{-5}$~eV, and the total energy within $10^{-6}$~eV. The
core electron states were treated full-relativistically, whereas
the valence states were treated both, non-relativistic and
full-relativistically, i.e. including the spin-orbit interaction,
besides the Darwin and mass-velocity corrections.

In addition, we have also used the DFT-LDA, with the plane-wave
description of the wavefunctions, and the pseudopotential method
({\it Abinit} code~ \cite{gonze02}) for the evaluation of the
phonon contributions to the real part of the dielectric function
($\varepsilon_1^{latt}(w)$), as will be discussed later. With this
approach, we are able to unveil and estimate the total value of
the static dielectric constant of ZrO$_2$ in the form
$\varepsilon_{1}(0)$ = $\varepsilon_1^{elect}(0)$ +
$\varepsilon_1^{latt}(0)$, for the three
 phases cubic, tetragonal, and monoclinic.

\section{Results and Discussions}

\subsection{Structural properties}

The cubic ZrO$_2$ phase (c-ZrO$_2$) belongs to the space group
$Fm3m$ with crystalline structure of fluorite, i.e. face-centered
cubic lattice with three atoms at the basis, being one Zr atom at
the position (0,0,0) and the oxygen atoms at the position ($1/4$,
$1/4$, $1/4$), for which the BZ is the usual 14-faces polyhedra.
We have obtained  through a total energy minimization process, a
lattice constant value of the cubic phase $a=5.139$ \AA, slightly
larger than that calculated by Kr\'alik {\it et al} \cite{Kralik},
using the pseudopotential method and the local density
approximation (LDA), and closer to the experimental value of
$5.09$ \AA \cite{Stefanovich}. The value obtained by us confirms
that the GGA approximation leads to lattice constant values which
are slightly overestimated in relation to the experimental value.

The tetragonal phase (t-ZrO$_2$) is described by the $P42/nmc$
group. As shown by Zhao and Vanderbilt \cite{Zhao}, the t-ZrO$_2$
phase is obtained from the cubic one, by properly performing
alternate displacements $\Delta c$ of the oxygen atoms in the
cubic structure. Our optimization process consisted in minimizing
the total energy and forces, until "converged" $\Delta c$, $c$,
and $a$ values have been reached. The obtained values were
$a=5.10$ \AA, $c=5.23$ \AA, and the internal parameter d$_{z}$ =
${{\Delta c}/c}=0.05$, close to the experimental ones
\cite{Stefanovich} ($a=5.05$ \AA, $c=5.18$ \AA, and d$_{z}$ =
0.0574), as well as close to the theoretical values reported by
Kr\'alik {\it at al.} \cite{Kralik} ($a=5.04$ \AA~and $c=5.10$
\AA~ and d$_{z}$ = 0.0423).

For the monoclinic phase, which is described by the $P21/c$ group,
we have evaluated the total energy for several lattice constant
values, in order to have a good description of the equilibrium
properties for the bulk m-ZrO$_{2}$, in which case, for sake of
reducing computational time, we use the {\it {Abinit}}
code~\cite{gonze02}. The results were fitted by the Murnaghan
equation of state, and we have obtained, for the monoclinic
structure, $a= 5.12$ \AA, $b= 5.16$ \AA, $c= 5.33$ \AA, and
$\theta$ = 99.6 degrees.  These results are in accordance
 with the available experimental \cite{Stefanovich}($a=
5.15$ \AA, $b= 5.21$ \AA, $c= 5.31$ \AA, and $\theta$ = 99.23
degrees), and theoretical results \cite{Walter}($a= 5.20$ \AA, $b=
5.25$ \AA, $c= 5.41$ \AA, and $\theta$ = 99.60 degrees). We have
used the later ones in the band structure calculations performed
with Wien2k code, as well as to calculate effective mass values
and optical properties for the monoclinic phase.

\subsection{Electronic properties}

Table I presents the band gap energies, together with the
valence-to-conduction band transition symmetry, for all  three
phases of ZrO$_2$. For the cubic phase, the band gap is indirect
at $X \longrightarrow \Gamma$, with an energy of $3.09$~eV when a
non-relativistic calculation is considered, and with energy of
$3.30$~eV, if the calculations are performed by taking into
account the full relativistic effects, i.e. the
scalar-relativistic and spin-orbit effects. The direct band gap,
at $\Gamma $, is of $3.61$ ($3.80$)~eV within a non-relativistic
(full-relativistic) calculation.  Our values  for the  band gaps
are very close to those obtained by Kr\'alik {\it et al.}
\cite{Kralik}, in which a pseudopotential LDA method was used. We
recall that, the theoretical values for the band gap energy are
smaller when compared with the experimental one, due to the
well-known underestimation of conduction band states energies in
{\it ab initio} calculations which are performed within DFT.
Experimental values, as reported by French {\it et al.}
\cite{French}, are found in the range of $6.1$ to $7.08$~eV.

Due to the spin-orbit interaction, we observe a spin-orbit (so)
splitting energy of the valence band top ($v$), at the
$\Gamma$-point, which for the cubic phase is
$\Delta^{v}_{so}(\Gamma)= 69$~meV. There are no experimental data
for $\Delta_{so}$ in ZrO$_2$ reported so far.

For the tetragonal phase, the energy transitions at $X
\longrightarrow \Gamma$, $Z \longrightarrow \Gamma$, and $\Gamma
\longrightarrow \Gamma$ are very near. An indirect gap of $3.80$
($4.01$)~eV, and a direct one of $\sim 3.9$ ($\sim 4.1$)~eV, were
obtained from non-relativistic (full-relativistic) calculations. A
value of $\Delta^{v}_{so}(\Gamma)= 9$~meV has been observed for
the spin-orbit splitting energy of the valence band top, at
$\Gamma$, for t-ZrO$_2$. The band gap energy values, obtained
through the full-relativistic calculations, are well compared with
those obtained by Kr\'alik {\it et al.} \cite{Kralik}. The
experimental values for t-ZrO$_2$ lie in the range of $5.8$ to
$6.6$~eV, according to the results reported by French {\it et al.}
\cite{French}.

 For the monoclinic
phase, we have obtained an indirect gap of $\sim 3.4$~eV  ($\sim
3.6$)~eV in a non-relativistic (full-relativistic) calculation,
and a direct band gap of $3.50$~eV ($3.64$)~eV, so much close to
the value of the indirect one, that it is difficult to assure
whether the gap is direct or indirect. As one goes from the cubic
to tetragonal, and then to the monoclinic structural phase, the
symmetry lowering is responsible for the removal of the
degeneracy, hence to the appearance of so many energy bands in the
monoclinic case.

Figure 1 shows the non-relativistic results obtained for the band
structure, along high symmetry directions of the BZ, and the total
density of states (TDOS) for the three phases of ZrO$_2$. The {\it
main} character of the peaks in the TDOS is also emphasized. The
results for band structures, as obtained from the
full-relativistic calculations are presented in Fig. 2. In the
upper valence part of the band structure, the O(2p)-related states
are predominant, while in its lower part, the O(2s)- derived
states are the most relevant, independently of inclusion of the
relativistic corrections, for all three phases. The Zr(4p) states
appear rather deep in the valence band (-26 eV) and, due to the
spin-orbit coupling are split into a group at -26 and a group at
-28 eV, corresponding to j=3/2 and j=1/2 manifolds, respectively.

\subsection{Carrier effective masses}

The valence- and conduction-band effective masses for the cubic,
tetragonal, and monoclinic phases of ZrO$_2$, at relevant symmetry
points of the BZ, are presented in Table II for several
directions. Full- and non-relativistic (see the values in
parenthesis) results are shown. In order to remain within the
region of validity of the parabolic approximation, the effective
masses were calculated by considering the energy curves within a
small region of radius $0.5 \%$  centered on the extremum of
interest.

As we observe, in both the valence and conduction bands, the
carrier effective masses are very anisotropic, i.e., show a
relevant dependence on the $\vec{k}$ direction. Moreover,
remarkable differences arise when the relativistic effects are
taken into account. In particular, for the cubic phase,
relativistic effects are seen to be most important in the $\Gamma
- X$ direction, for holes and electrons, while  these effects are
more relevant in the $\Gamma - Z$ direction, for the case of
holes, for the tetragonal phase, and in the $\Gamma-Y$ direction,
for electrons and holes, for the monoclinic phase. So far, there
are no reported experimental data in the literature of carrier
effective masses in ZrO$_2$. This represents the first report of
effective mass values, as calculated for all phases of ZrO$_2$, of
fundamental importance to the modelling of advanced devices based
on this oxide.

\subsection{Optical properties}

The electronic contributions to the real
($\varepsilon_{1}(\omega)$) and imaginary
($\varepsilon_{2}(\omega)$) parts of the complex dielectric
function, $\varepsilon(\omega)=\varepsilon_{1}(\omega)+
\varepsilon_{2}(\omega)$, can be obtained  from the band structure
directly through the Wien2k code \cite{wien2k}, $\varepsilon_{1}$
 from $\varepsilon_{2}$, and vice-versa, using the
Kramers-Kronig relations \cite{Cardona Book}. Once is known the
imaginary part $\varepsilon_{2}(\omega)$, from the FLAPW
electronic structure calculations, the real part
$\varepsilon_{1}(\omega)$ is then given by \cite{Cardona Book}

\begin{center}
\begin{equation}
\label{Eq1} \varepsilon_1(\omega)= 1 +
(2/{\pi}){\int_{0}^{\infty}} d\omega^{\prime}{{\omega^{\prime 2}
\varepsilon_2 (\omega^{\prime}) \over {\omega^{\prime 2} -
\omega^2}}}.
\end{equation}
\end{center}

Other optical properties, such as the reflectivity R($\omega$) and
the complex refraction index n($\omega$) can also be obtained. We
took into account that energy gaps are under evaluated by DFT
\cite{GGA}, by adopting the following procedure: the dielectric
function, and all the related optical properties, were obtained by
performing an  upward rigid shift, to the experimental energy
value of the band gap \cite{French}, of all the conduction band
states. In general, for metals and conductor materials,  the
optical properties calculations demand a great number of
eigenvalues and eigenvectors, therefore, a study of convergence,
that considers the number of k-points used, is necessary. In the
case  of the insulating oxide ZrO$_2$ studied here, we have
verified that no more than about $100$ symmetrized k-points were
sufficient.

Figure 3 shows the imaginary part of the dielectric function
(electronic contribution), for the three phases of ZrO$_2$,
considering energies till 30 eV. For the cubic phase, all the
diagonal components of $\varepsilon_2$ are identical
($\varepsilon_{xx} = \varepsilon_{yy}  = \varepsilon_{zz}$),
whereas  the off-diagonal components are zero. The energy spectrum
was {\it adjusted} by shifting it upwards, to the band gap
experimental value for the monoclinic phase, $E_{g} = 5.83$ eV
\cite{French}. For tetragonal and monoclinic phases, the
adjustment was taken  by considering the same experimental value
as for the monoclinic phase, once this later is the observed phase
at ambient temperature. The function $\varepsilon_2$, for
t-ZrO$_2$, has all off-diagonal components equal to zero, but
symmetry assures that $\varepsilon_{xx} = \varepsilon_{yy} \neq
\varepsilon_{zz}$. For the monoclinic phase, the components that
are different from zero are $\varepsilon_{xx}$,
$\varepsilon_{yy}$, $\varepsilon_{zz}$, and $\varepsilon_{xz}$.

In order to avoid confusion, results for both full-relativistic
(frel) and non-relativistic (nrel) calculations are shown in Fig.3
only for the cubic phase. Relative changes in the values of the
dielectric function, due to relativistic effects amount to $ 5
\%$, at maximum, for all cases. For the tetragonal and monoclinic
phases, the presented results derive from full-relativistic
calculations. A comparison of the theoretical findings with the
experimental result of Camagni {\it et al.} \cite{Camagni}, as
derived from reflectivity data, shows better agreement for the
monoclinic phase for which two more pronounced peaks occur in the
theoretical curve, at $7.5$~eV and $\sim 11.5$~eV.

The real part of the complex dielectric function, $\varepsilon_1$,
is presented in Fig. 4, for the three  phases of ZrO$_2$. As for
the case of $\varepsilon_2$ (shown in Fig.3), we show results for
$\varepsilon_1$ as obtained from full-relativistic and
non-relativistic calculations only for c-ZrO$_2$; for the other
two phases, tetragonal and monoclinic, the results depicted in
parts (b) and (c) of Fig.4 correspond to full-relativistic
calculations.

In order to compare the theoretical result with   experiment, we
show in Fig.5 the calculated $xx$-component of the real part of
the dielectric function, $\varepsilon_{1xx}$, for m-ZrO$_2$, till
energies of $30$~eV, together with the data reported by Camagni et
al. \cite{Camagni} for two samples of m-ZrO$_2$ stabilized with
$12\%$ and $24\%$ of Yttria. Excellent accordance is seen between
theory and experiment.

In Fig. 6, we show the calculated reflectivity (R) for the three
phases of ZrO$_2$, together with the experimental results, as
extracted from French {\it et al.} \cite{French} and  from Camagni
{\it et al.}~\cite{Camagni}. In the former work, the authors have
considered three different samples of ZrO$_2$, in cubic,
tetragonal, and monoclinic phases. These experimental results are
referred  as Exp.[2] in Fig.6. The data in the figure, referred as
Exp.[1], correspond to m-ZrO$_2$ single crystals stabilized with
$12\%$ of Yttria \cite{Camagni}. Again, a good agreement between
theoretical and experimental results is observed for the
monoclinic phase.

Figure 7 shows the behavior of the refraction index, as function
of energy, for all three phases of ZrO$_2$, as derived from
full-relativistic calculations. For the cubic phase, we also
included the result obtained from a non-relativistic calculation,
for comparison. In Table III are depicted the values of the
refraction index, at two different energies, for the tetragonal
phase. They compare well with the data reported by French {\it et
al.}, as obtained from VUV spectroscopy measurements
\cite{French}.

\subsection{Phonon Contribution to the Dielectric Function}

In order to realistically describe the dielectric function and, in
particular, to evaluate the static dielectric constant,
$\varepsilon_{1}(0)$, we have also to consider  the phonon
contribution to the low-frequency region of the
frequency-dependent dielectric function. The enhancement, observed
in the measured values of the ZrO$_2$ dielectric constant, is due
to the role played by the optical phonons. We have used the {\it
{Abinit}} code~\cite{gonze02} to unveil this phonon contribution
to the dielectric functions of the three phases. With this
program, we calculated the equations of state, the phonon
frequencies and eigendisplacements for the structures (not shown
in this work). In this case, we adopted the Troullier-Martins
pseudopotentials in the calculations~\cite{fuchs99}, and the
phonon dynamics were obtained by means of the adiabatic
perturbation density functional theory. According to the available
data, and previous theoretical results, for the phonon frequencies
[14], our results are in complete agreement.

From the obtained results, we have calculated the frequency
dependence of the real part of the dielectric constant (as
described by Rignanese {\em et al.}~\cite{rignanese04} and Gonze
{\em et al.}~\cite{gonze97}) for the monoclinic, tetragonal, and
cubic phases.

In Fig.~8 we show the results for $\varepsilon_{1}^{(latt)}(w)$,
for m-ZrO$_2$, along the $x$, $y$, and $z$ axis, together with a
fitting by means of the Lorentz-Drude model, i.e.

\begin{equation}
\label{Lorentz-Drude} \varepsilon_1^{(latt)}(\omega)=
\sum_{i=1}^{n} {{{\omega_{pi}^{2}(\omega_{TOi}^2 - \omega^2)}\over
{[(\omega_{TOi}^{2} - \omega^2)^2 + \gamma_{i}^2 \omega^2]}}},
\end{equation}

\noindent where  $n$ is the total number of oscillators, each one
with the transversal optical (TO) phonon frequency $\omega_{TOi}$,
$\omega_{pi}$ is the plasmon frequency for the $i^{th}$
oscillator, and $\gamma_{i}$ is its damping parameter, which is
directly related to sample size. The Drude-Lorentz curves are a
guide for the eye, and show that our calculated data follow this
model.  In the case of the $x$-component of
$\varepsilon_{1}^{(latt)}(w)$, we included four oscillators, and
the obtaining fitting parameters were $\omega_{p1}$ = 541.83
cm$^{-1}$ and $\gamma_{1}$ = 0.05 cm$^{-1}$ for the TO mode at 709
cm$^{-1}$, $\omega_{p2}$ = 635.54 cm$^{-1}$ and $\gamma_{2}$ =
0.05 cm$^{-1}$ for the TO mode at 485.1 cm$^{-1}$, $\omega_{p3}$ =
1052.34 cm$^{-1}$ and $\gamma_{3}$ = 0.05 cm$^{-1}$ for the TO
mode at 316 cm$^{-1}$, and $\omega_{p4}$ = 493.07 cm$^{-1}$ and
$\gamma_{4}$ = 0.05 cm$^{-1}$ for the TO mode at 220 cm$^{-1}$.

To describe the frequency dependence of the dielectric constant
parallel to the $y$ axis, given by the Lorentz-Drude model, we
needed five oscillators. The obtained fitting parameters were
$\omega_{p1}$ = 517.81 cm$^{-1}$ and $\gamma_{1}$ = 0.05 cm$^{-1}$
for the TO oscillator at 576 cm$^{-1}$, $\omega_{p2}$ = 493.44
cm$^{-1}$ and $\gamma_{2}$ = 0.05 cm$^{-1}$ for the TO oscillator
at 473 cm$^{-1}$, $\omega_{p3}$ = 997.54 cm$^{-1}$ and
$\gamma_{3}$ = 0.05 cm$^{-1}$ for the TO oscillator at 413
cm$^{-1}$, $\omega_{p4}$ = 576.96 cm$^{-1}$ and $\gamma_{4}$ =
0.05 cm$^{-1}$ for the TO oscillator at 354 cm$^{-1}$, and
$\omega_{p5}$ = 465.39 cm$^{-1}$ and $\gamma_{5}$ = 0.05 cm$^{-1}$
for the TO oscillator at 247 cm$^{-1}$.

Finally, we have also to include five oscillators, in order to
describe the frequency dependence of the dielectric constant
parallel to the $z$ axis. The obtained fitting parameters  were
$\omega_{p1}$ = 147.44 cm$^{-1}$ and $\gamma_{1}$ = 1.22 cm$^{-1}$
for the TO mode at 485.1 cm$^{-1}$, $\omega_{p2}$ = 353.93
cm$^{-1}$ and $\gamma_{2}$ = 0.81 cm$^{-1}$ for the TO mode at 399
cm$^{-1}$, $\omega_{p3}$ = 1029.0 cm$^{-1}$ and $\gamma_{3}$ =
2.18 cm$^{-1}$ for the TO mode at 354 cm$^{-1}$, $\omega_{p4}$ =
513.52 cm$^{-1}$ and $\gamma_{4}$ = 11.1 cm$^{-1}$ for the TO mode
at 316 cm$^{-1}$, and $\omega_{p5}$ = 327.06 cm$^{-1}$ and
$\gamma_{5}$ = 0.0001 cm$^{-1}$ for the TO mode at 220 cm$^{-1}$.

The calculated static dielectric tensor, due to lattice
contribution, for the monoclinic phase of ZrO$_2$ is

\begin{equation}\label{matrixphonon}
\left( {\begin{array}{*{20}c}
  24.12 & 0 & 1.77 \\
0 & 21.20 & 0 \\
1.77 & 0 & 17.02 \\
  \end{array}} \right).
\end{equation}

\noindent Therefore, using $\varepsilon_{1}(0)$ =
$\varepsilon_{1}^{(electr)}(0) + \varepsilon_{1}^{(latt)}(0)$, and
using the $\varepsilon_{1}^{(electr)}(0)$ from Figure~4(c) we get

\begin{equation}\label{matrixtotal}
\left( {\begin{array}{*{20}c}
   29.52 & 0 & 1.83 \\
0 & 26.20 & 0 \\
1.83 & 0 & 21.92 \\
  \end{array}} \right).
\end{equation}

\noindent The average value of the static dielectric constant, for
the monoclinic phase, is then given by
$\overline{\varepsilon}_{1}(0)$ $\sim$ 25.9, which agrees fairly
well with reported experimental values of $\sim$ 25~
[Ref.\cite{WilkI}].

It is worth mentioning that, we have also evaluated the dielectric
constants for t- and m-ZrO$_2$, by using the same approach
described above. The calculated phonon contribution for the
tetragonal phase, added to the corresponding electronic part, lead
to an average value of the static dielectric constant of
$\overline{\varepsilon}_{1}(0)=55.8$, a value which is as high as
that obtained for the cubic phase, $44.4$. Details of these
calculations will be published elsewhere \cite{Horacio phonons}.
These findings are another indication that the measured dielectric
constant for the ZrO$_{2}$ is correctly described by considering
the oxide in its monoclinic phase.

\section{Conclusions}

We have performed an extended study, based on {\it {ab initio}}
calculations, of the structural, electronic, and optical
properties for the cubic, tetragonal, and monoclinic crystalline
phases of ZrO$_2$, which takes into account scalar relativistic
and spin-orbit contributions to the electronic structure, as well
as the phonon contributions to the low-frequency regime of the
dielectric function. For the band structure calculations, we used
the DFT-GGA approach by means of the FLAPW method. Due to the
spin-orbit interaction, we observe a spin-orbit (so) splitting
energy of the top of the valence band ($v$) at the $\Gamma$-point,
which for the cubic phase is $\Delta^{v}_{so}(\Gamma)= 69$~meV,
while for the tetragonal phase is $\Delta^{v}_{so}(\Gamma)=
9$~meV. So far, there are no experimental data reported for
$\Delta_{so}$ in ZrO$_2$.

The carrier effective masses are shown to be highly anisotropic,
with the relativistic corrections playing an important role; this
is of special interest when modelling advanced ZrO$_2$-based
devices. In addition we report,  with the relativistic
corrections, an evaluation of a spin-orbit effective mass in the
valence band, as well as the heavy- and light-hole (electron)
effective masses at the valence (conduction) band, in particular
for cubic  ZrO$_2$.

The results obtained for the real and imaginary parts of the
dielectric function,  for the reflectivity, and for the refraction
index, show good agreement with the available experimental
results. In order to obtain the static dielectric constant of
ZrO$_2$, we added to the electronic part, the optical phonons
contribution, which leads to values of $\varepsilon_{1}(0)\simeq
29.5, 26.2, 21.9$, along the $xx, yy$, and $zz$ directions,
respectively, for the monoclinic phase, in excellent accordance
with experiment.

Relativistic effects, taking also into account  the spin-orbit
interaction, are demonstrated to be important for a better
evaluation of the effective mass values, and in the detailed
structure of the frequency dependent complex dielectric function.
We were able to unveil and estimate a more realistic total value
of the static dielectric constant of ZrO$_2$, in the form
$\varepsilon_{1}(0)$ = $\varepsilon_1^{elect}(0)$ +
$\varepsilon_1^{latt}(0)$, yielding an average value of $\sim$
25.9, in good agreement with reported experimental values of
$\sim$ 25.

\newpage

\acknowledgements

We acknowledge support from the Brazilian funding agencies, CNPq,
FAPESP, and FAPEMIG. One of the authors, SCPR, would like to
acknowledge the scholarship from the project CT-ENERG/CNPq (Grant
No. 503.570/03-6).

\newpage

\begin{figure}[ht]
\includegraphics[scale=0.6,clip=true,angle=-90.0]{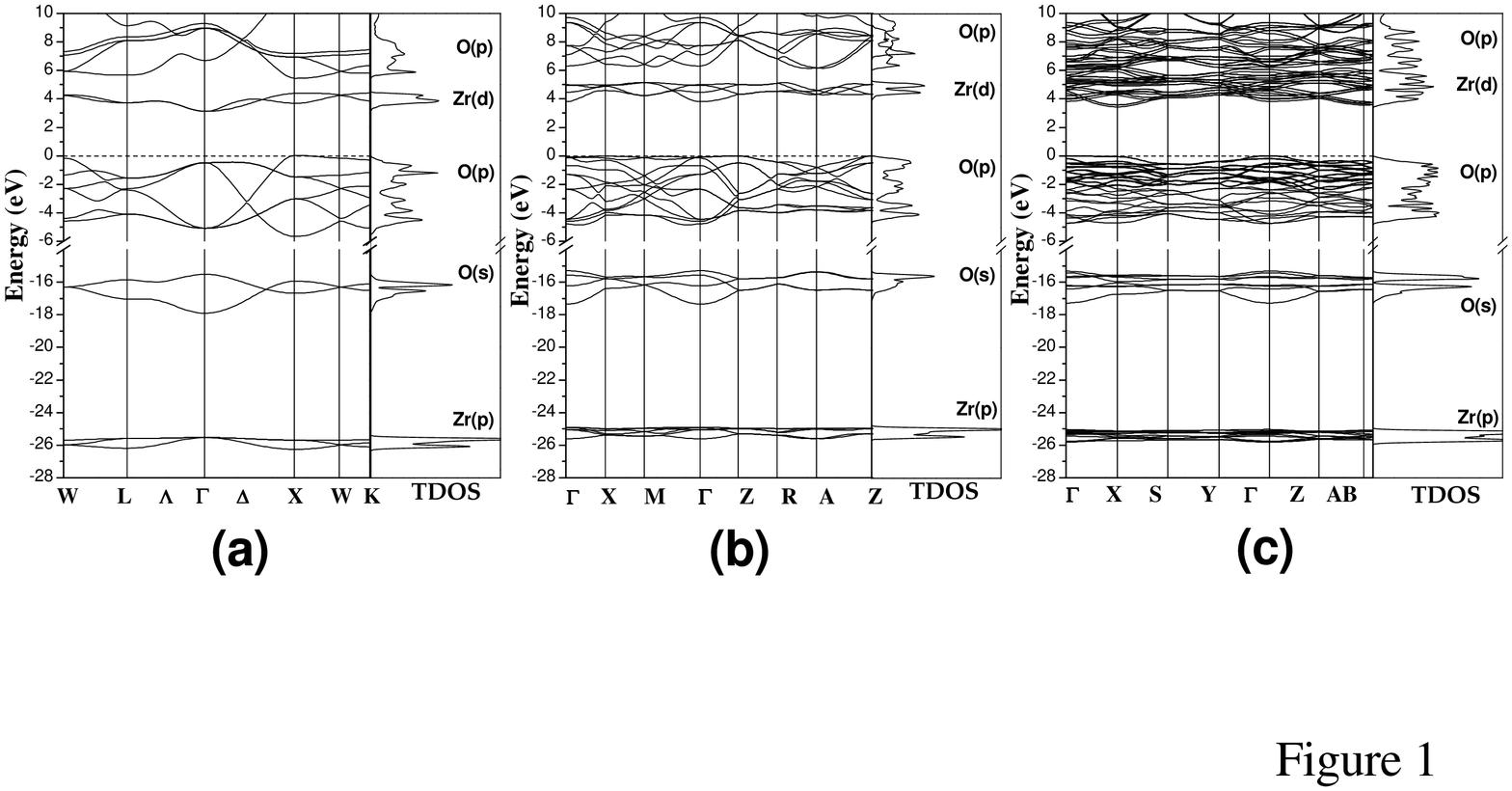}
\caption{Band structures, along high symmetry axis of the BZ, and
total density of states (TDOS) for (a) cubic, (b) tetragonal, and
(c) monoclinic phases of ZrO$_2$, as obtained from
non-relativistic calculations. The energy zero was taken at the
valence band maximum, shown by a dashed horizontal line. The main
character of the peaks in the TDOS is emphasized.} \label{bandNon}
\end{figure}

\begin{figure}[ht]
\includegraphics[scale=0.6,clip=true,angle=-90.0]{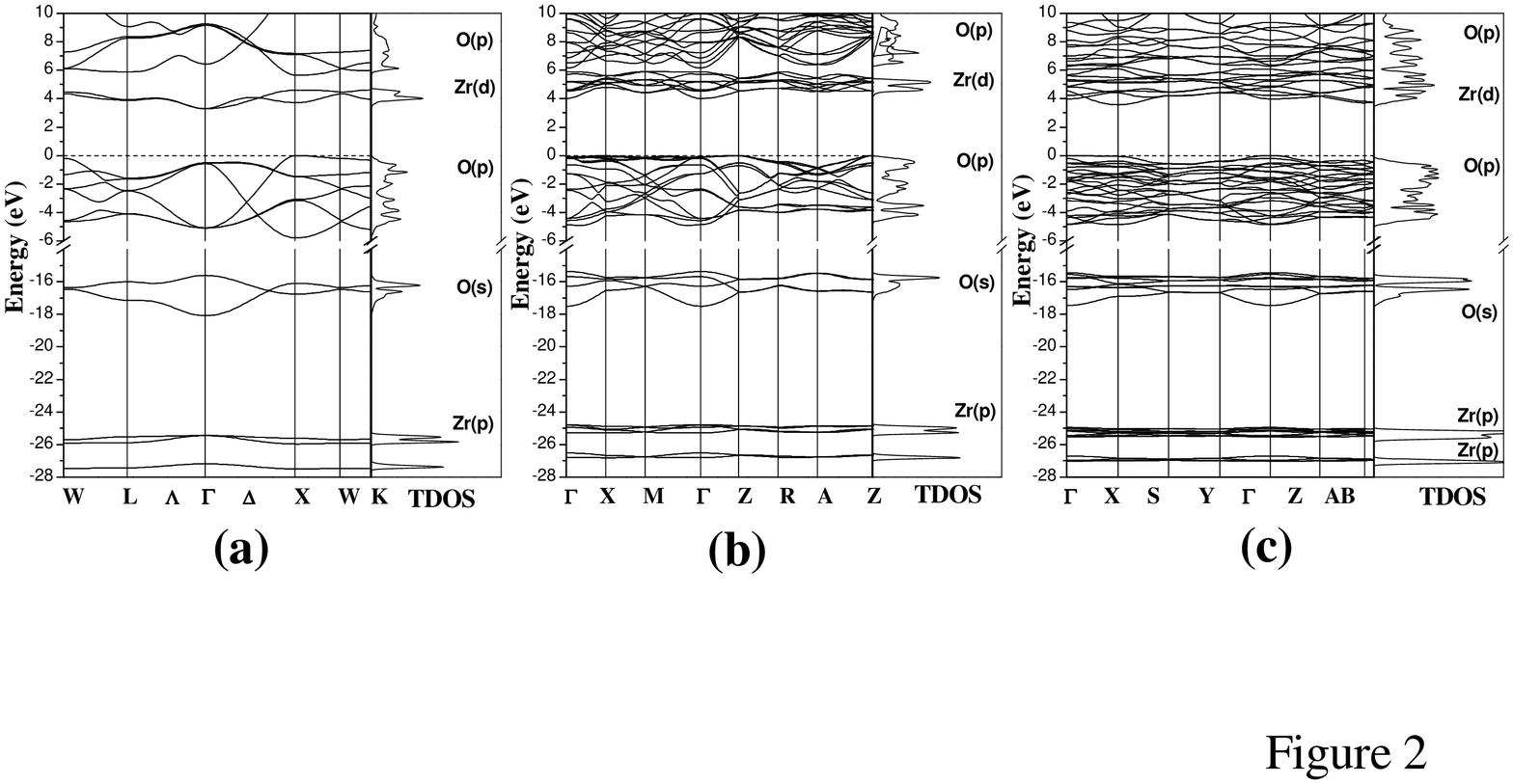}
\caption{Band structures, along high symmetry axis of the BZ, and
total density of states (TDOS) for (a) cubic, (b) tetragonal, and
(c) monoclinic phases of ZrO$_2$, as obtained from
full-relativistic calculations. The energy zero was taken at the
valence band maximum, shown by a dashed horizontal line. The main
character of the peaks in the TDOS is emphasized.}
\label{bandFull}
\end{figure}

\begin{figure}[ht]
\includegraphics[scale=0.6,clip=true,angle= 270.0]{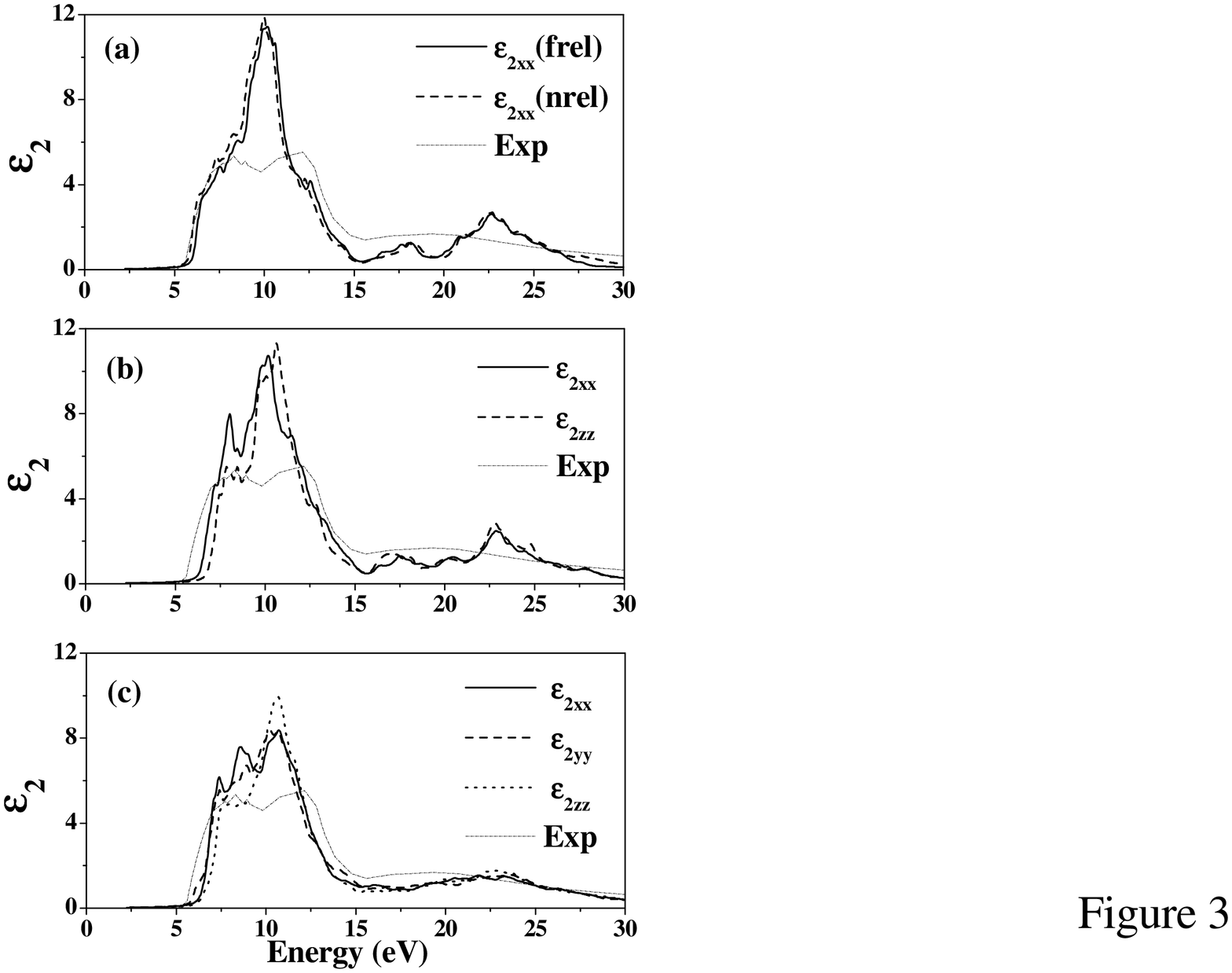}
\caption{Imaginary part of the dielectric function,
$\varepsilon_{2}$, versus energy, for the (a) cubic, (b)
tetragonal, and (c) monoclinic phases of ZrO$_2$, as obtained from
full-relativistic calculations. Only in (a), results as obtained
from full-relativistic calculations ($frel$) and non-relativistic
calculations ($nrel$) are shown, for comparison. The theoretical
curves were adjusted by performing a rigid shift upwards in energy
to the experimental gap value for the monoclinic phase
$E_{g}^{exp}=5.83$~eV [Ref. \cite{French}]. } \label{epsilon2}
\end{figure}

\begin{figure}[ht]
\includegraphics[scale=0.6,clip=true,angle= 270.0]{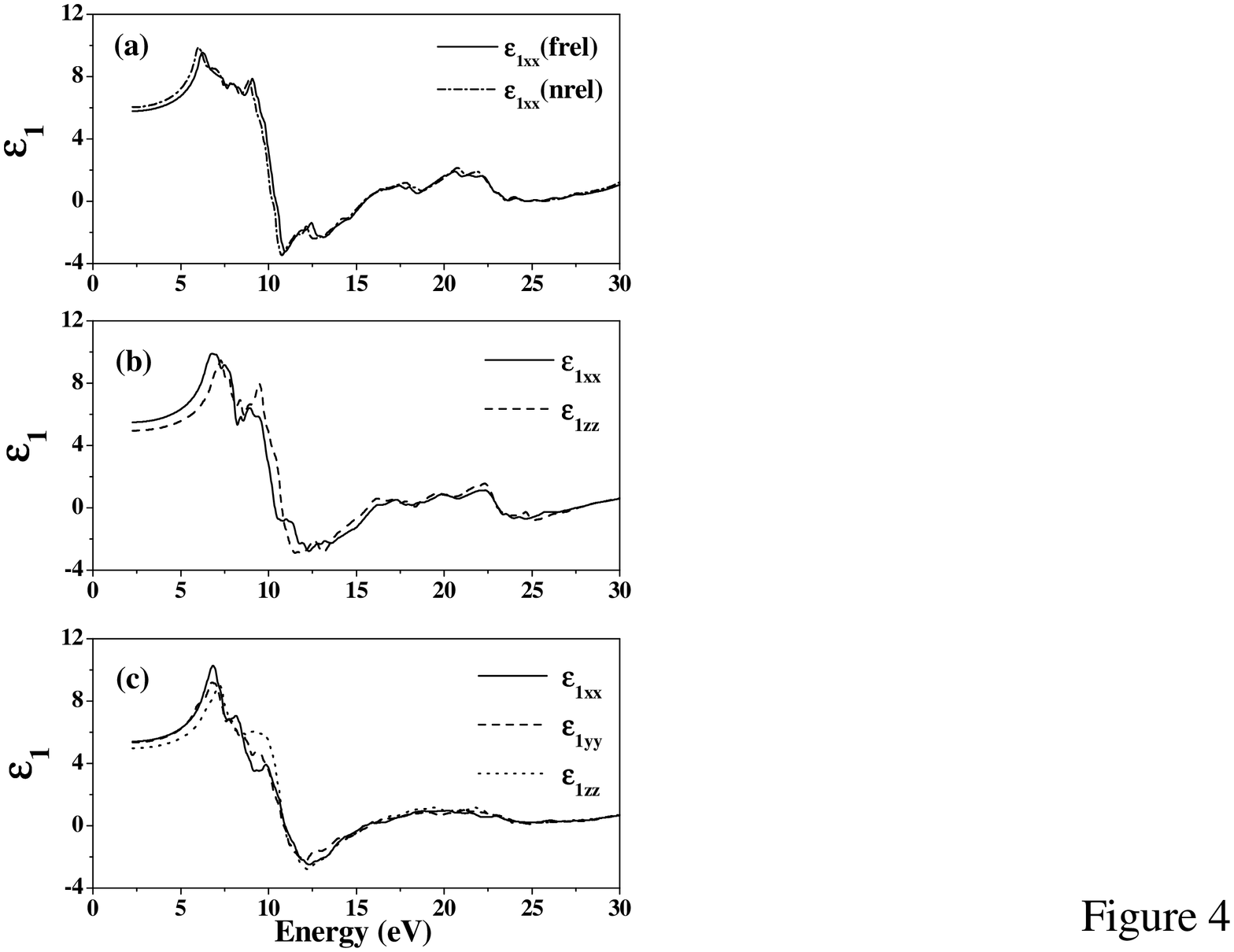}
\caption{Real part of the dielectric function, $\varepsilon_{1}$,
versus energy, for the (a) cubic, (b) tetragonal, and (c)
monoclinic phases of ZrO$_2$, as obtained from full-relativistic
calculations. Only in (a), results as obtained from
full-relativistic calculations ($frel$) and non-relativistic
calculations ($nrel$) are shown, for comparison. The theoretical
curves were adjusted by performing a rigid shift upwards in energy
to the experimental gap value for the monoclinic phase
$E_{g}^{exp}=5.83$~eV [Ref. \cite{French}]. } \label{epsilon1}
\end{figure}

\begin{figure}[ht]
\includegraphics[scale=0.6,clip=true,angle= 270.0]{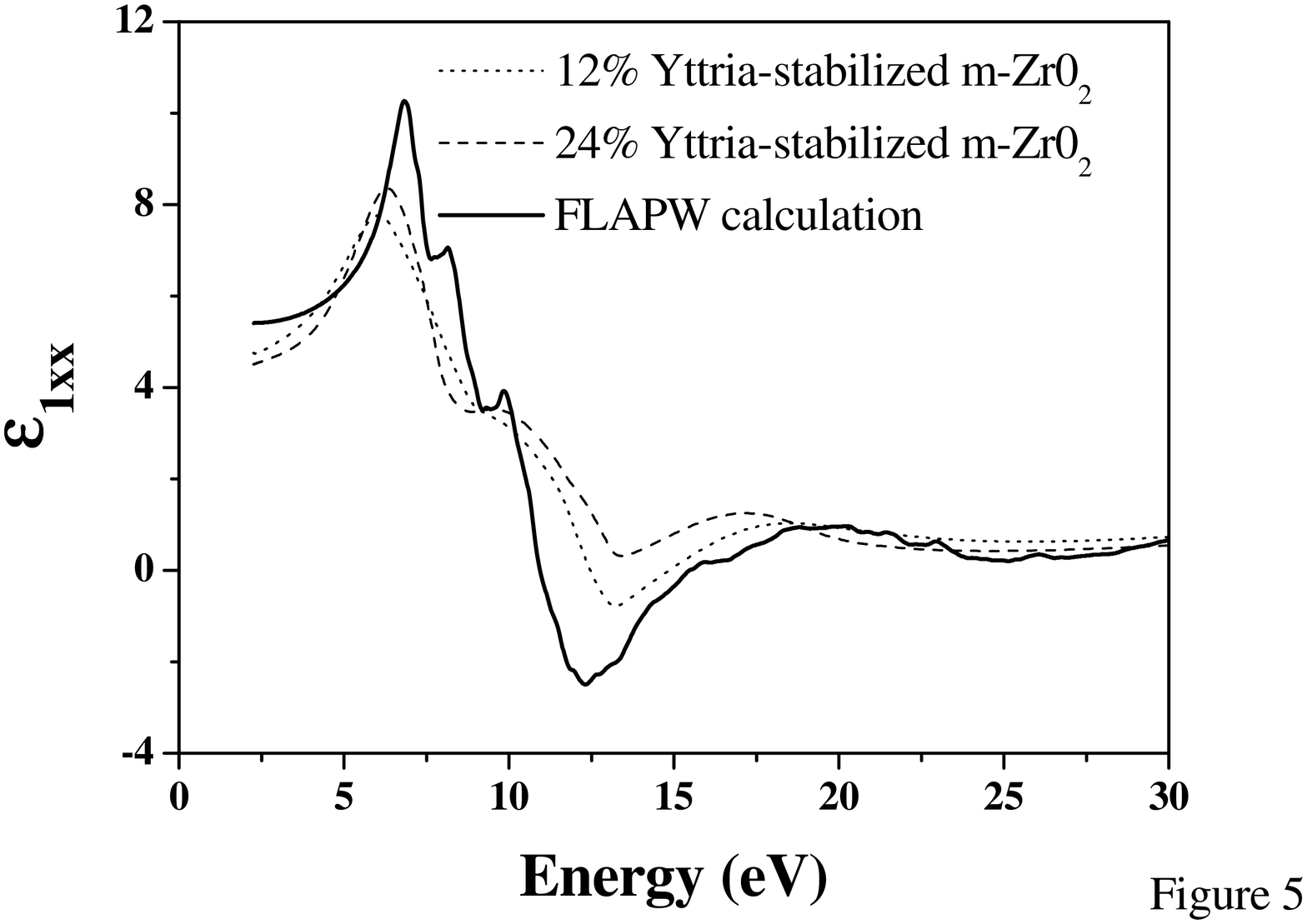}
\caption{$xx$-component of the real part of the dielectric
function, $\varepsilon_{1xx}$, for monoclinic ZrO$_2$, as obtained
from full-relativistic FLAPW calculations, together with the
experimental results as derived from reflectivity data, extracted
from Ref.\cite{Camagni} for two yttria-stabilized samples.}
\label{epsilon1exp}
\end{figure}

\begin{figure}[ht]
\includegraphics[scale=0.6,clip=true]{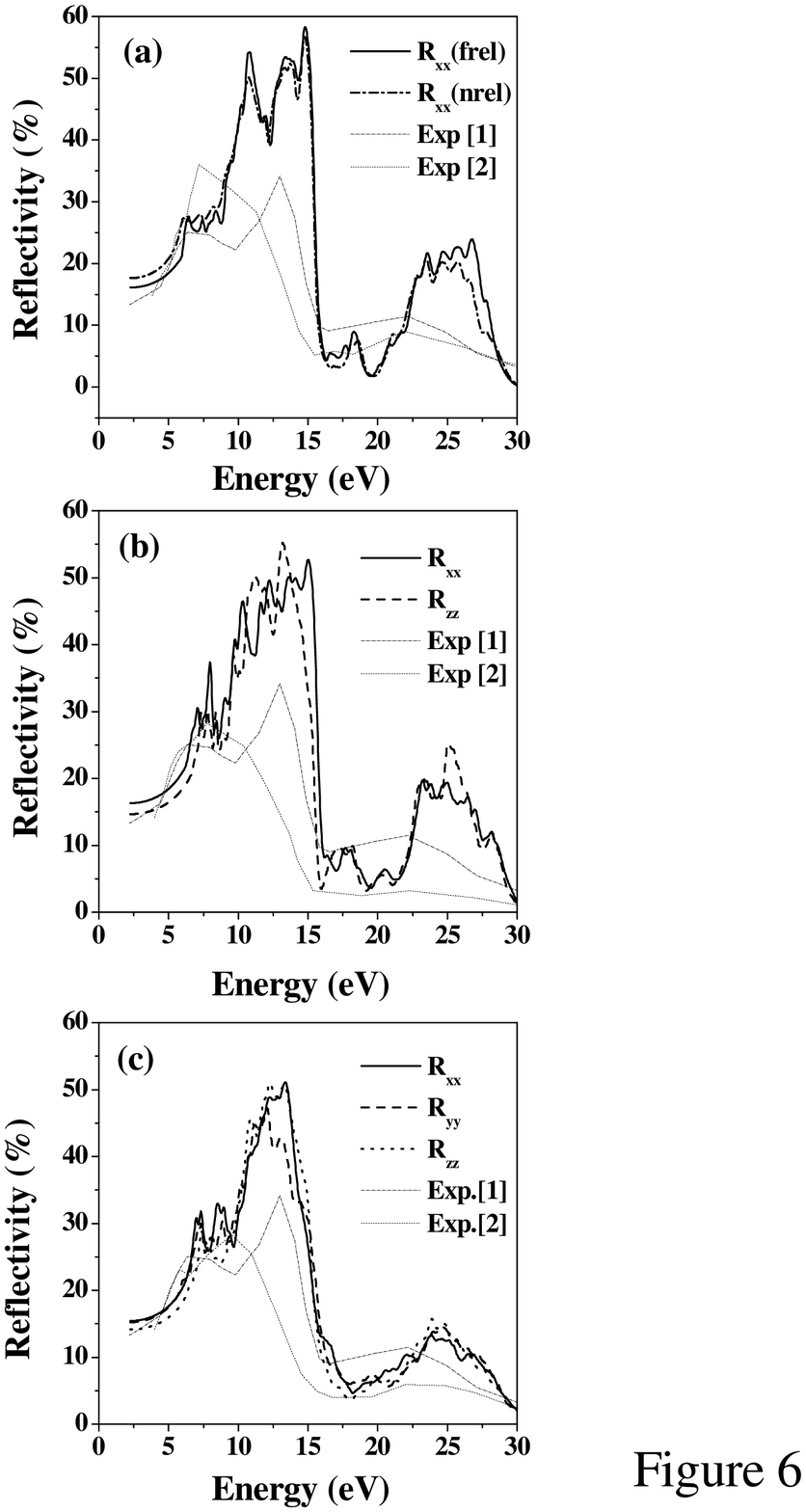}
\caption{Reflectivity (in $\%$) versus energy, for the (a) cubic,
(b) tetragonal, and (c) monoclinic phases of ZrO$_2$, as obtained
from full-relativistic calculations. Only in (a), results as
obtained from full-relativistic calculations ($frel$) and
non-relativistic calculations ($nrel$) are shown, for comparison.
Also shown are the experimental curves, as extracted from French
{\it et al.} \cite{French} (Exp.[1]) and from Camagni {\it et
al.}\cite{Camagni} (Exp.[2]).} \label{reflectivity}
\end{figure}

\begin{figure}[ht]
\includegraphics[scale=0.6,clip=true]{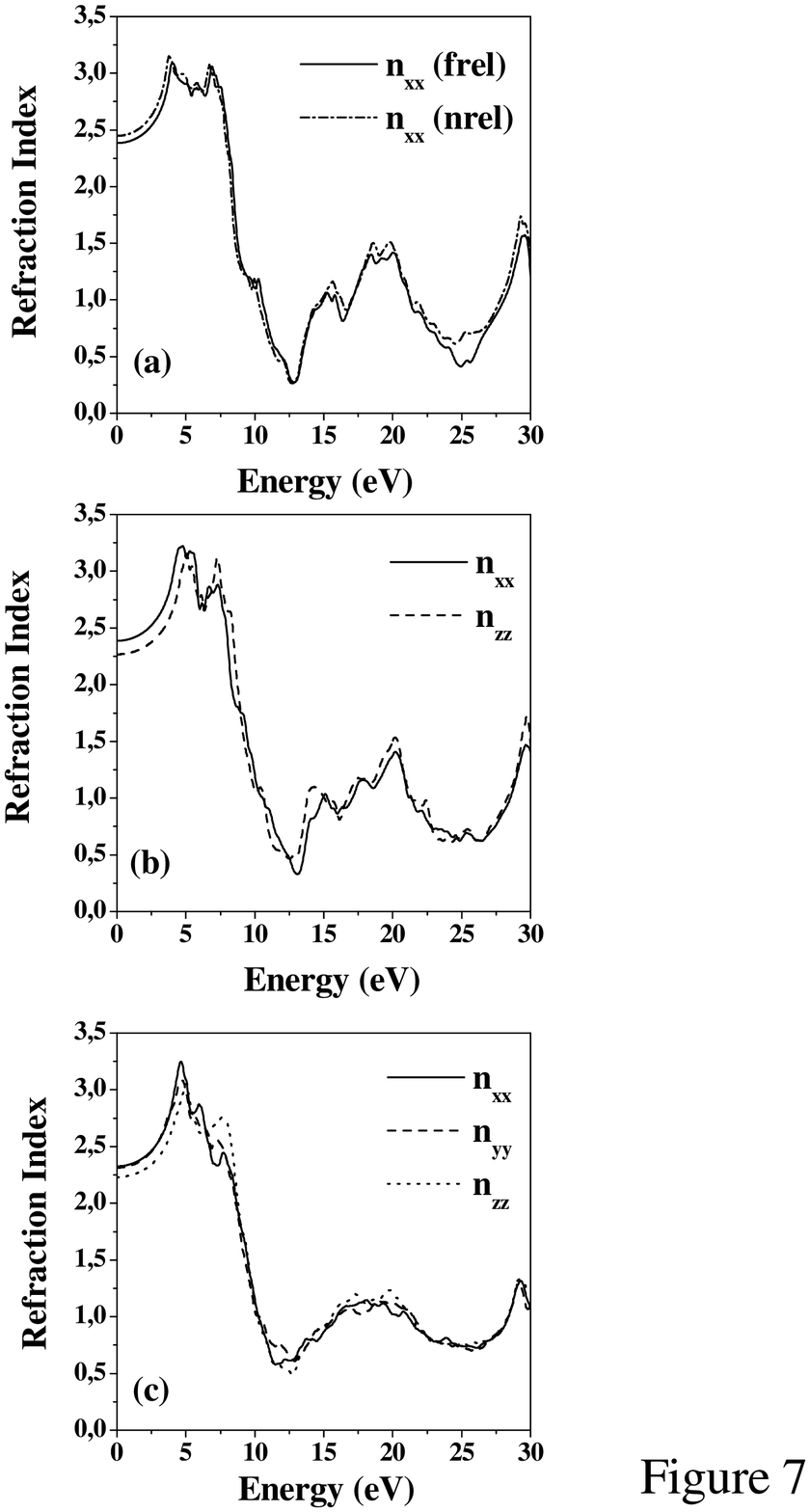}
\caption{Refraction index versus  energy for the (a) cubic, (b)
tetragonal, and
 (c) monoclinic phases of ZrO$_2$, as obtained from full-relativistic
calculations. Only in (a), results as obtained from
full-relativistic calculations ($frel$) and non-relativistic
calculations ($nrel$) are shown, for comparison.}
\label{refractionindex}
\end{figure}

\begin{figure}
\includegraphics[scale=0.6,clip=true]{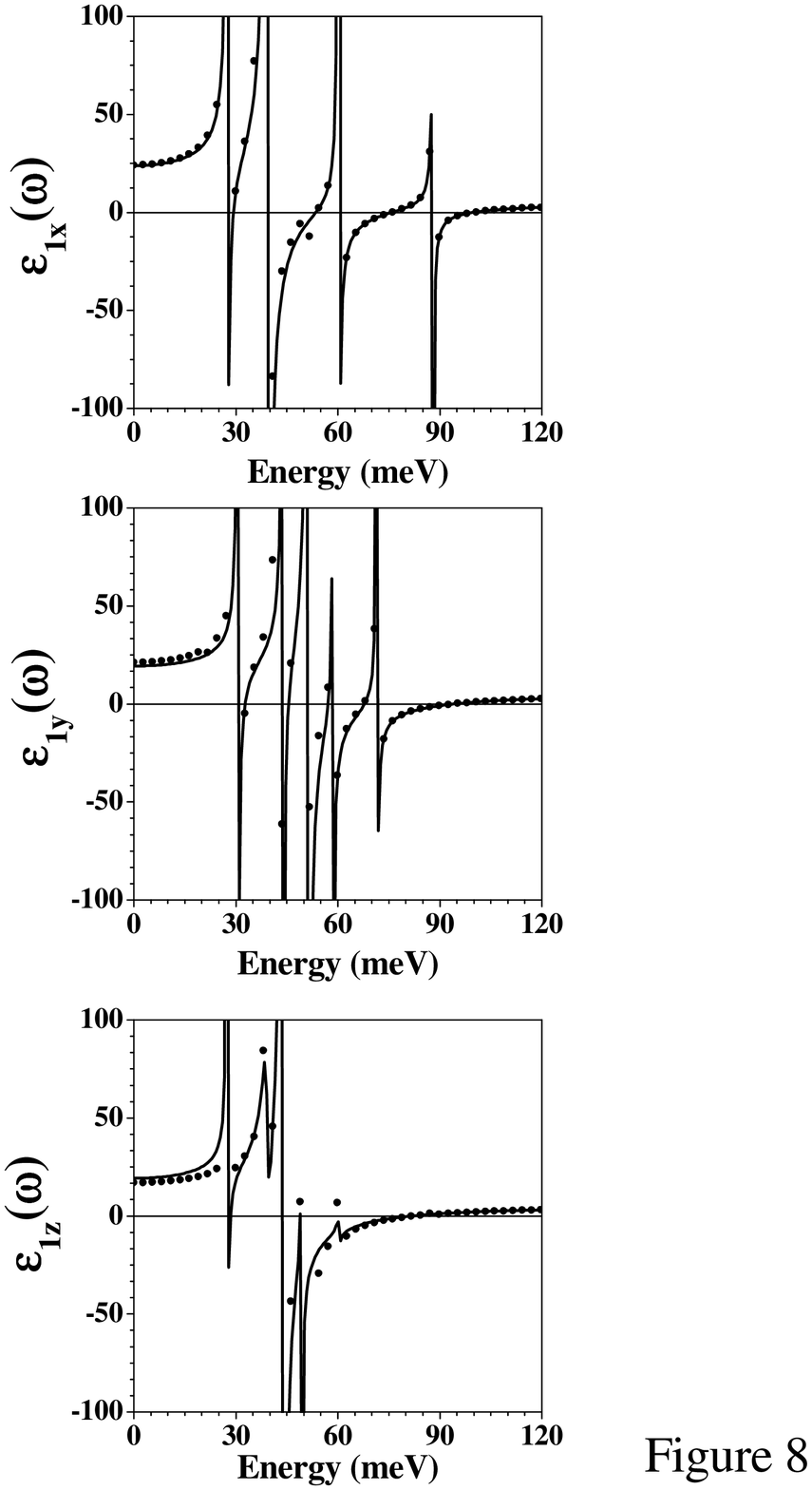}
\caption{Phonon contribution (full circles) to the low-frequency
dependence of the real part of the dielectric function,
$\varepsilon_{1}^{latt}(w)$, as obtained from {\it ab initio}
calculations, parallel to the $x$, $y$, and $z$ axis, for
monoclinic ZrO$_2$. The solid line corresponds to a fitting by
means of the Lorentz-Drude expression, and was used as a guide to
the eye.} \label{epsilon1phonons}
\end{figure}

\newpage


\begin{table}
\caption{Band gap energies, $E_g$ (in eV), and symmetry of the
valence-to-conduction band transition, obtained from the
non-relativistic ($E_g^{nrel}$) and full-relativistic
($E_g^{frel}$) calculations for the cubic, tetragonal, and
monoclinic  phases of ZrO$_2$.}
\begin{center}
\begin{tabular}{cccc}
\hline \hline  phase & valence-to-conduction band transition
symmetry~~ & $E_g^{nrel}$ ~~  & $E_g^{frel}$  \\
 \hline \hline
    cubic~~ &  $X \longrightarrow \Gamma$ & 3.09  &  3.30  \\
 ~~ &  $\Gamma \longrightarrow \Gamma$ & 3.61  &  3.80  \\
       ~~ &  $X \longrightarrow X$ & 3.65  &  3.72  \\ \hline
      tetragonal~~ &  $Z \longrightarrow \Gamma$ & 3.80  &  4.01  \\
    ~~ &  $X \longrightarrow \Gamma $ & 3.83  &  4.04  \\
           ~~ &  $\Gamma \longrightarrow \Gamma $ & 3.88  &  4.09  \\
            \hline
       monoclinic~~  &  $\Gamma \longrightarrow X$ & 3.44  &  3.58  \\
     ~~  &  $\Gamma \longrightarrow \Gamma$ & 3.82  &  3.98
\\
              ~~  &  $X \longrightarrow X$ & 3.50  &  3.64  \\
\hline \hline
\end{tabular}
\end{center}
\label{gaps}
\end{table}

\newpage


\begin{table}
\caption{Valence and conduction band effective masses (in units of
the rest free electron mass, $m_0$), at relevant symmetry points
of the BZ, for the cubic (c), tetragonal (t), and monoclinic (m)
phases of ZrO$_2$, as obtained from full-relativistic band
structure calculations. In parenthesis, are shown the effective
mass values as derived from non-relativistic calculations.
$m^{*}_{h}$, $m^{*}_e$, $m^{*}_{he}$, $m^{*}_{le}$,
 $m^{*}_{hh}$, $m^{*}_{lh}$, and $m^{*}_{so}$ stand for, respectively, hole,
electron, heavy-electron, light-electron, heavy-hole, light-hole,
and split-off hole effective masses. }
\begin{center}
\begin{tabular}{cccc}
\hline \hline  phase& direction & valence~ & conduction
\\
 \hline \hline
  c   &  $\Gamma \longrightarrow L$ & $m^{*}_{so}=0.26$;$m^{*}_{lh}=0.23$(0.23);$m^{*}_{hh}=0.28$(0.27)
     &  ~~$m^{*}_{he}= 0.28(0.26)$;$m^{*}_{le}= 0.27$(0.26)  \\
   &  $\Gamma \longrightarrow X$& $m^{*}_{so}=0.77$;$m^{*}_{lh}=0.36$(0.24);$m^{*}_{hh}=3.84$(4.23)
   &  ~~$m^{*}_{he}=2.09$(2.06);$m^{*}_{le}= 0.51$(0.48) \\
        &  $X \longrightarrow \Gamma$ & $m^{*}_{h}=0.28$(0.27) &  $m^{*}_{e}=1.77$(1.97) \\
       ~~ &  $X \longrightarrow W$ & $m^{*}_{h}=3.38$(3.29) &
$m^{*}_{e}=1.17$(1.33)  \\ \hline
       t &  M $\longrightarrow \Gamma$~~ & $m^{*}_h=0.47$(0.47) &  ...(...)  \\
  &  $\Gamma \longrightarrow Z $~~ & $m^{*}_h=5.31$(6.68)  &
$m^{*}_{e}=2.32$(2.32)   \\
~~ &  $Z \longrightarrow R $ & $m^{*}_{h}=0.83$ (0.88) & ... (...)  \\
\hline
        m  &  $X \longrightarrow \Gamma$ ~~& ...(...)  & $m^{*}_{e}=1.16$(1.12) \\
       &  $ X \longrightarrow S$~~ & $m^{*}_{h}=1.09$(1.06) &  $m^{*}_{e}=2.32$(2.32)  \\
              ~~  &  $\Gamma \longrightarrow Y$~~ & $m^{*}_{h}=2.55$(3.08)
& $m^{*}_{e}=1.39$(0.75)  \\
              ~~  &  $\Gamma \longrightarrow Z$~~ & $m^{*}_{h}=1.49$(1.42)
&  $m^{*}_{e}=4.08$(4.18)  \\
\hline \hline
\end{tabular}
\end{center}
\label{masses}
\end{table}

\newpage


\begin{table}
\caption{Values for the refraction index, at certain energies, as
obtained from non-relativistic ($n^{nrel}$) and full-relativistic
($n^{frel}$) calculations, for tetragonal  ZrO$_2$. For
comparison, the experimental results as observed from VUV
spectroscopy measurements, and extracted from Ref. \cite{French},
are shown in the last column ($n_{exp.}$). $n_{xx}$ and $n_{zz}$
stand,
 respectively, for perpendicular and parallel refraction index to the
axis along $c$, one of the lattice constants in the tetragonal
lattice.}
\begin{center}
\begin{tabular}{cccccc}
\hline \hline  Energy (eV)~~ & $n_{xx}^{nrel}$~~ & $n_{xx}^{frel}$
~~
& $n_{zz}^{nrel}$  ~~ &$n_{zz}^{frel}$ & $n_{exp.}$ \\
 \hline \hline
  1.96~~ &  2.52 ~~ & 2.47 ~~ &  2.38~~  & 2.33~~ & 2.192  \\
 2.54~~ &  2.59~~ & 2.53~~  &  2.43~~  & 2.38~~& 2.208  \\
\hline \hline
\end{tabular}
\end{center}
\label{refraction index}
\end{table}




\newpage


\end{document}